\documentclass{emulateapj}
\usepackage{apjfonts}
\usepackage{epsf}
\usepackage{psfig}
\begin{document}

\slugcomment{Submitted to ApJL}
\shortauthors{Miller, Miller, \& Reynolds}
\shorttitle{Spin and Supernovae}

\title{The Angular Momenta of Neutron Stars and Black Holes as a Window on Supernovae}

\author{J.~M.~Miller\altaffilmark{1},
        M.~C.~Miller\altaffilmark{2},
        C.~S.~Reynolds\altaffilmark{2}}

\altaffiltext{1}{Department of Astronomy, University of Michigan, 500
Church Street, Ann Arbor, MI 48109-1042, jonmm@umich.edu}

\altaffiltext{2}{Department of Astronomy, University of Maryland,
  College Park, MD, 20742}


\label{firstpage}

\begin{abstract}
It is now clear that a subset of supernovae display evidence for jets
and are observed as gamma-ray bursts.  The angular momentum
distribution of massive stellar endpoints provides a rare means of
constraining the nature of the central engine in core-collapse
explosions.  Unlike supermassive black holes, the spin of stellar-mass
black holes in X-ray binary systems is little affected by accretion,
and accurately reflects the spin set at birth.  A modest number of
stellar-mass black hole angular momenta have now been measured using
two independent X-ray spectroscopic techniques.  In contrast,
rotation-powered pulsars spin-down over time, via magnetic braking,
but a modest number of natal spin periods have now been estimated.
For both canonical and extreme neutron star parameters, statistical
tests strongly suggest that the angular momentum distributions of
black holes and neutron stars are markedly different.  Within the
context of prevalent models for core-collapse supernovae, the angular
momentum distributions are consistent with black holes typically being
produced in GRB-like supernovae with jets, and with neutron stars
typically being produced in supernovae with too little angular
momentum to produce jets via magnetohydrodynamic processes.  It is
possible that neutron stars are imbued with high spin
initially, and rapidly spun-down shortly after the supernova event,
but the available mechanisms may be inconsistent with some observed
pulsar properties.
\end{abstract}

\section{Introduction}
The angular momentum or spin period of a rotating star is easily
defined and understood.  To make a consistent comparison with black
holes, however, the dimensionless angular momentum must be used.  This
quantity is referred to as the ``spin parameter'' or ``spin'' of a
black hole, and it is given by $a = {\rm cJ}/{\rm GM}^{2}$, where $-1
\leq a \leq 1$ (where J is the angular momentum, G is Newton's
gravitational constant, and M is the mass of the black hole).
Interactions between the disk and hole likely enforce a limit of $a
\leq 0.998$ (Thorne 1974), assuming zero torque at the innermost disk
edge.  For likely neutron star radii and masses, rotation at the
break-up frequency would give a spin parameter of $a \simeq 0.7$.

Efforts to measure and constrain cosmic acceleration, for instance,
are aided by the ability to rely on samples of sources rather than
individual measurements, and by the fact of independent measurement
techniques.  The case is similar in the young enterprise of black hole
spin measurement: independent measurement techniques are being
exploited, and small samples have been compiled.  Modeling thermal
continuum emission from the accretion disk provides one means of
measuring black hole spin (see, e.g., Shafee et al.\ 2006, McClintock
et al.\ 2006).  Modeling the dynamical broadening of emission lines
excited in the disk by external irradiation (disk reflection) provides
a second, independent means of measuring black hole spin (see, e.g.,
Miller et al.\ 2002, 2009; Reis et al.\ 2009; for a recent review, see
Miller 2007).  Both methods rely on the premise that the innermost
edge of the accretion disk is sensitive to the innermost stable
circular orbit (ISCO; see, e.g. Bardeen, Press, Teukolsky 1972), which
is a complex but monotonic function of the black hole spin parameter.

A black hole with zero spin must accrete half of its initial mass to
achieve a spin of $a = 0.84$, and considerably more mass to achieve
even higher spins (Bardeen 1970).  In an X-ray binary, then, accretion
does not strongly affect the spin of the hole: a low mass companion
cannot donate enough mass, and a high mass companion cannot donate
enough mass in its short lifetime.  As a result, the spin of
stellar-mass black holes reflects the value imparted at birth.  In
contrast, the spin of neutron stars can change dramatically.  Magnetic
braking can act to spin-down a neutron star,
while accretion in an X-ray binary can act to greatly increase the
spin frequency.  However, in a small but growing number of isolated,
rotation-powered pulsars, it has recently been possible to estimate
the natal spin period of the star.  This requires a measurement of the
magnetic dipole braking index, and/or a knowledge of the age of the
supernova remnant in which the neutron star was formed. 

Detections of supernovae-like spectra in the afterglow of gamma-ray
bursts clearly associate long-duration gamma-ray bursts (GRBs) with a
subset of core-collapse supernovae (Bloom et al.\ 1999; Stanek et
al.\ 2003).  The ``collapsar'' model (Woosley 1993; MacFadyen \&
Woosley 1999) is widely accepted as the standard model for
long-duration GRBs.  The jets predicted in these supernova models are
in broad agreement with evidence of beaming (Frail et al.\ 2001).  The
spin parameter of stellar endpoints can provide a rare window into the
central engine of supernovae and GRBs, and may be important in
understanding why a subset of supernovae display jets.  In this work, we
exploit the small but growing number of stellar-mass black hole spin
measurements and natal neutron star spin periods, and explore
potential consequences for prevalent supernova models.

\section{Data Selection}
A set of nine estimated natal spin periods for isolated,
rotation-powered pulsars is given in Table 7 of Faucher-Giguere \&
Kaspi (2006).  These sources and their natal spin periods are treated
in detail in this analysis.  In the two cases where an initial period
is given as a lower limit, we have conservatively adopted the limiting
value (this will give higher dimensionless angular momenta).


\begin{deluxetable}{lll}
\tablecaption{Black Hole Angular Momenta}
\tablehead{
\colhead{Source}  & \colhead{cJ/GM$^{2}$} &  \colhead{cJ/GM$^{2}$} \cr
\colhead{~} & \colhead{(reflection)} & \colhead{(continuum)}
}     
\startdata
M33 X-7           &           --         &           0.77(5)$^{f}$ \\

LMC X-1           &           --         &            0.92(6)$^{g}$ \\

A 0620$-$00       &            --         &            0.12(19)$^{h}$ \\

4U 1543$-$475      &           0.3(1)$^{a}$    &             0.80(5)$^{i}$ \\

XTE J1550$-$564      &         0.76(1)$^{a}$     &           -- \\

XTE J1650$-$500      &         0.79(1)$^{a}$     &           -- \\

XTE J1652$-$453      &         0.45(2)$^{b}$     &           --  \\

GRO J1655$-$40       &         0.98(1)$^{a}$     &           0.70(5)$^{i}$ \\

GX 339$-$4           &         0.94(2)$^{a}$     &           -- \\

SAX J1711.6$-$3808    &        0.6(3)$^{a}$      &           -- \\

XTE J1752$-$223         &      0.55(11)$^{c}$    &           -- \\

Swift J1753.5$-$0127    &      0.76(13)$^{d}$    &           -- \\

XTE J1908$+$094        &       0.75(9)$^{a}$      &          -- \\

GRS 1915$+$105         &       0.98(1)$^{e}$       &         0.99(1)$^{j}$ \\

Cygnus X-1            &        0.05(1)$^{a}$       &         -- \\

\enddata
\tablecomments{Measured values of black hole spin parameters are given above.  The errors are statistical errors on the last significant digit.  $^{a}$ Miller et al.\ (2009).  $^{b}$ Hiemstra et al.\ (2010).  $^{c}$ Reis et al.\ (2010).  $^{d}$ Reis et al.\ (2009).  $^{e}$ Blum et al.\ (2009).  $^{f}$ Liu et al.\ (2008).  Gou et al.\ (2009).   $^{h}$ Gou et al.\ (2010). $^{i}$ Shafee et al.\ (2006).  $^{j}$ McClintock et al.\ (2006).  }
\end{deluxetable}


The measurement of black hole spins is in its infancy, and so we
imposed some modest quality metrics in selecting black hole data.  We
required that all spin data must be based on statistical fits wherein
a trial spectral model was folded through an instrument response and
evaluated against an observed spectrum using a goodness-of-fit
statistic.  Owing to the lack of errors {\it and} goodness-of-fit
statistics, then, we have excluded early spin estimates by Zhang, Cui,
\& Chen (1997).  Blum et al.\ (2009) report two spin values for
GRS~1915$+$105 based on fits to the disk reflection spectrum obtained
with {\it Suzaku}.  The higher value of $a = 0.98\pm 0.01$ is used in
this work, as it derives from fits made to a spectrum spanning a
much broader energy range.  A lower value for the spin of GRS 1915$+$105
has also been reported based on fits to the disk continuum (Middelton
et al.\ 2006).  The high value listed in Table 1 is likely more robust
as it was derived by giving extra weight to observations wherein a
standard Novikov-Thorne (1973) accretion disk is likely to hold (McClintock
et al. 2006).  A quality cut is automatically imposed on spin
measurements made using thermal continuum emission from the accretion
disk, in that the mass and distance to a source must be
well-determined if fits are to yield meaningful constraints.

Table 1 details the black hole dimensionless angular momenta used in
this work.  In five of the black hole systems listed (XTE J1652$-$453,
SAX J1711.6$-$3808, XTE J1752$-$223, Swift J1753.5$-$0127, and XTE
J1908$+$094), strong dynamical constraints requiring a black hole
primary have not yet been obtained through radial velocity techniques.
However, myriad phenomena observed from of these sources are
consistent with dynamically-constrainted systems, and it is extremely
likely that these X-ray binaries harbor black holes (for a review of
characteristic properties, see Remillard \& McClintock 2006).

\section{Analysis}
Black hole angular momenta are measured using relativistic
spectroscopic models; no additional calculations are requied to make
use of these data.  In the case of neutron stars, it is necessary to
calculate the angular momentum of each star using its mass and spin
period.  If the equation of state of ultra-dense matter were known,
the moment of inertia could be calculated by integrating the
prescription for how mass varies with radius.  At present, the correct
equation of state is unknown, and many candidates exist (see Lattimer
\& Prakash 2006).  It is worth noting that some soft equations of
state have recently been ruled-out with the discovery of an especially
massive neutron star ($M = 1.97\pm 0.04~M_{\odot}$; Demorest et
al.\ 2010).  In order not to bias our analysis in favor of any
particular equation of state, we have simply approximated the moment
of inertia as ${\rm I} = (2/5){\rm M}{\rm R}^{2}$ as per a sphere of
uniform density.

Dimensionless angular momenta were calculated using the natal periods
given in Faucher-Giguere \& Kaspi (2006), for different combinations
of neutron star mass and radius.  One set of dimensionless angular
momenta were calculated assuming canonical parameters: ${\rm R} =
10$~km and ${\rm M}_{\rm NS} = 1.4~{\rm M}_{\odot}$.  Extreme
dimensionless angular momenta result from assuming a larger radius for
a given mass.  All of the equations of state treated in Lattimer \& Prakash
(2006) predict radii less than or equal to 15~km.  A set of extreme
angular momenta were therefore derived assuming ${\rm R} = 15$~km and
${\rm M}_{\rm NS} = 1.4~{\rm M}_{\odot}$.

The mean and median values of the dimensionless angular momentum
distributions are given in Table 2, and the distributions are shown in
Figure 1.  The black hole distributions are consistent with
moderately high values: $a_{\rm mean} = 0.66$ for spins derived using
disk reflection spectra, and $a_{\rm mean} = 0.72$ for spins derived
using the disk continuum.  These values stand in marked contrast to
those derived for the neutron star samples.  For neutron stars where
the natal spin period has been estimated, even extreme stellar
parameters only work to yield $a_{\rm mean} = 0.029$.  

\begin{deluxetable}{lll}
\tablecaption{Distribution Properties}
\tablehead{
\colhead{Sample~~~~~~~~~~}  & \colhead{$a_{\rm mean}$~~~~~~~~} &  \colhead{$a_{\rm median}$~~~~~~~~} \cr
}     
\startdata
BH (reflection) & 0.66  & 0.76 \\

BH (continuum) & 0.72   & 0.80 \\

NS (natal; $1.4~{\rm M}_{\odot}$, R=15~km) & 0.029 & 0.017 \\

NS (natal; $1.4~{\rm M}_{\odot}$, R=10~km) & 0.018 & 0.007 \\

\enddata
\end{deluxetable}


The sample of black hole spin parameters derived using disk reflection
fits is small (12 measurements), but twice as large as the present
sample derived using the disk continuum.  A Gaussian fit to the
reflection-derived distribution gives $a_{\rm cent} = 0.71$ and
$\sigma = 0.26$.  A fit to the smaller distribution of
continuum-derived spin parameters gives $a_{\rm cent} = 0.81$ and
$\sigma = 0.06$ (the width should be viewed as a lower limit as the
fit was largely insensitive to the lowest spin value).  Although
likely aided by the small sample sizes currently available, the spin
distributions found using these independent methods are formally
consistent.  The sample of natal neutron star periods is also small,
with only nine estimates.  Assuming the extreme stellar parameters, a
Gaussian fit to this distribution gives $a_{\rm cent} = 2.1\times
10^{-2}$, and $\sigma = 1.9\times 10^{-2}$.  Assuming canonical
stellar parameters, a Gaussian fit gives $a_{\rm cent} = 9.5\times
10^{-3}$, and $\sigma = 8.4\times 10^{-3}$.  

Table 3 lists the results of running two-sided Kolmogorov-Smirnov
tests on different angular momentum distributions.  This statistical
test measures the probability that two samples are drawn from the same
parent distribution.  For the samples and assumed stellar parameters
considered in this work, the highest probability that the neutron star
and black hole distributions are drawn from the same parent is just
$3.6\times 10^{-4}$.  Despite the fact that the sample of black hole
spin parameters and natal neutron star spin periods is small, it is
clear that they belong to distinct distributions.  Black holes are likely
born with high spin parameters, and neutron stars may be born with very
low spin parameters.

At the time of writing, the Australia Telescope National Facility
Pulsar Catalog (Manchester et al.\ 2005) contains 1875
rotation-powered pulsars.  The spins of the pulsars in this
compilation do not reflect birth conditions.  We simply note that no
pulsar in this sample has a dimensionless spin parameter in excess of
$a = 0.3$, neither for typical nor extreme stellar parameters, and the sample
is statistically distinct from the black hole distributions at
extremely high confidence.

It is difficult to assess the degree to which current black hole spin
measurements may be affected by systematic errors and skew
comparisons to the sample of neutron star angular momenta.  The thermal
continuum emission from stellar-mass black hole accretion disks can be
fit well with simple, two-parameter models that lack inner torque
conditions and provide no direct spin constraint.  New models that
include spin as a variable parameter are more complex, and rely on
accurate measures of the mass and distance of the black hole, as well
as reasonable estimates of the mass accretion rate and scattering in
the disk atmosphere (see, e.g., McClintock et al.\ 2006; Nowak et
al.\ 2008).  Spin constraints derived based on fits to disk reflection
spectra have primarily made use of relatively simple models which
assume a constant disk density (see, e.g., Miller et al.\ 2009).  A few
spectral fits have been made with more physical reflection models;
these fits appear to give consistent results (see, e.g., Reis et
al.\ 2008), but more work is needed. 

For three black holes in Table 1, spins have been measured using both
techniques, and the values do not agree in detail.  All of the spin
values and their associated errors were derived based on spectral
fitting that minimized a $\chi^{2}$ statistic.  The resulting
$\chi^{2}$ space can be complex, and using a 1$\sigma$ error to
evaluate the degree to which measurements disagree can give an
over-estimate (see Miller et al.\ 2009).  The measurements made for 4U
1543$-$475 are most strongly at odds.  In this case, the low
resolution and calibration of the gas proportional counter spectrum
may serve to under-estimate the extent of the iron line; a single
spectral bin could account for $\Delta(a) \simeq 0.1-0.2$.  The low
spin value for Cygnus X-1 may be another good example of a suspicious
value.  It derives from fits to the source in a ``low/hard'' state,
wherein the disk may not always extend to the ISCO (see, e.g. Esin,
McClintock, \& Narayan 1997).  Measurements of the inner disk radius based
on relativistic reflection modeling in states where the disk is more
certain to extend to the ISCO give values as small as $r_{in} = 2.2\pm
1.0~ GM/c^{2}$ (Miller et al.\ 2005), which translates into $a =
0.9^{+0.1}_{-0.2}$.

In two cases, then, potential systematic errors resulting from limited
spectral resolution and source phenomenology may have served to
under-estimate the degree to which black hole and neutron star angular
momentum distributions differ.  An uncertainty in basic accretion disk
physics, however, could potentially work in the opposite sense.  Both
techniques assume that the inner edge of an actual accretion disk is
truncated at the test-particle ISCO.  If real fluid accretion disks
push across the ISCO defined by test particle orbits, both spin
measurement techniques would give falsely high spin values.  New
numerical simulations aimed at addressing this specific issue appear
to justify the underlying assumption of both techniques (see, e.g.,
Shafee et al.\ 2008, Reynolds \& Fabian 2008).  However, this must be
continually revisited as numerical techniques and computing power
advance.

\section{Discussion and Conclusions}
The dimensionless angular momenta of stellar-mass black holes and
rotation-powered neutron stars are compared statistically.  Two
different methods of measuring black hole spin parameters are
considered, and both canonical and extreme stellar parameters were
considered for neutron stars with estimated natal spin periods.  A
strong statistical difference between the distributions signals that
black holes could typically born with dramatically higher dimensionless
angular momentum than neutron stars.  The observed dichotomy in their
dimensionless angular momenta implies that there may be important
differences in the core-collapse supernovae events that create these
stellar remnants.

The collapse of a $35~{\rm M}_{\odot}$ star with a $14~{\rm
  M}_{\odot}$ He core is examined in detail by MacFadyen \& Woosley
(1999).  Reasonable pre-collapse angular momentum values for the
eventual iron core are found to lead to the formation of a disk that
rapidly accretes onto a newly-formed black hole, quickly giving a spin
parameter of $a \simeq 0.9$.  Our results are consistent with
this prediction (see Tables 1 and 2, and Figure 1), lending
support to the overall model described by MacFadyen \& Woosley
(1999).  Of course, this model is the standard ``collapsar'' picture
for the production of gamma-ray bursts, and it is interesting to note
that the same disk required for driving the new black hole to a high
spin parameter is also responsible for producing a jet via MHD
processes.


\begin{deluxetable}{lll}
\tablecaption{Statistical Tests}
\tablehead{
\colhead{Sample~~~~~~~~}  & \colhead{Sample~~~~~~~~} &  \colhead{KS Probability} \cr
}     
\startdata

BH (reflection) & NS (natal; $1.4~{\rm M}_{\odot}$, 15~km)  & $9.3\times 10^{-5}$ \\

BH (reflection) & NS (natal; $1.4~{\rm M}_{\odot}$, 10~km)  & $1.4\times 10^{-5}$ \\

BH (continuum) & NS (natal; $1.4~{\rm M}_{\odot}$, 15~km)   & $3.6\times 10^{-4}$ \\

BH (continuum) & NS (natal; $1.4~{\rm M}_{\odot}$, 10~km)   & $3.6\times 10^{-4}$ \\

\enddata
\end{deluxetable}


Shibata \& Shapiro (2002) used simulations to understand the spin of
the black hole left by a uniformly rotating supermassive star, finding
$a \simeq 0.75$.  Shapiro \& Shibata (2002) later showed that the same
result can be derived analytically (also see Gammie, Shapiro, \&
McKinney 2004).  These treatments are primarily concerned with
extremely massive stars as candidate progenitors for the build-up of
supermassive black holes in the early universe.  Although our results
are consistent with these predictions as well, the detailed
models of MacFadyen \& Woosley (1999) may be more relevant to
the specific case of stellar-mass black holes.  

For both canonical and extreme stellar parameters, the sample of
natal neutron star spin periods strongly implies that these remnants
could be born with very low dimensionless 

\centerline{~\psfig{file=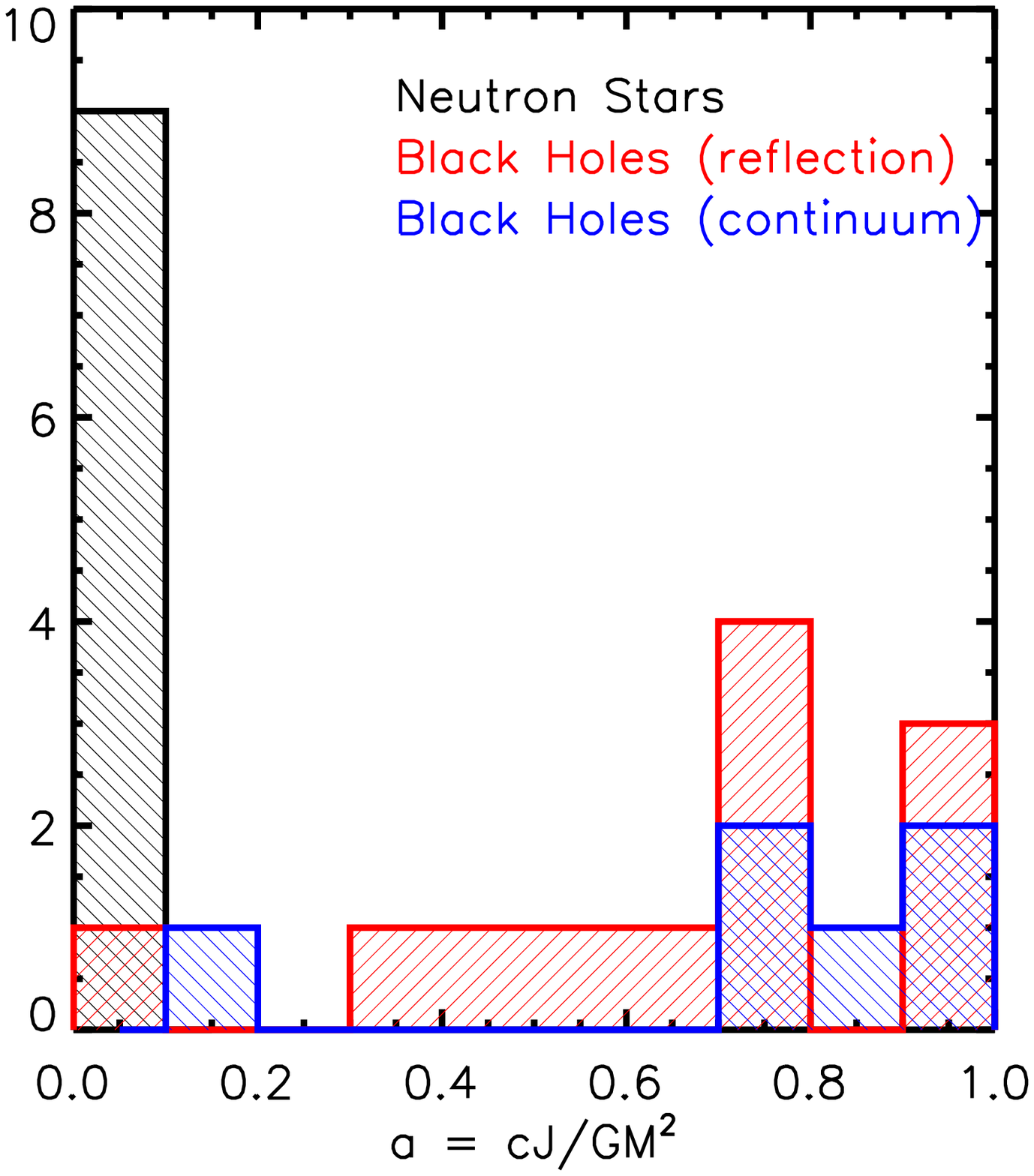,width=3.2in}~}
\figcaption[t]{\footnotesize The distribution
  of dimensionless angular momenta for neutron stars and stellar-mass
  black holes, compiled from recent measurements, is shown here.  The neutron star
  momenta were calculated using the subset of rotation-powered pulsars
  wherein natal spin periods have been estimated.  Stellar radii of
  15~km and masses of $1.4 {\rm M}_{\odot}$ were assumed in all cases in
  order to give the greatest possible angular momentum values.  A two-sided
  Kolmogorov-Smirnov test was used to evaluate the probability that
  neutron star and black hole spins were drawn from the same parent
  distribution.  A probability of $3.6\times 10^{-4}$ is found when
  comparing neutron star spins to black hole spins derived using the
  disk continuum.  The probability is $9.3\times 10^{-5}$ when using
  black hole spins derived using disk reflection.}
\medskip

\noindent angular momenta.  This stands in
stark contrast to some theoretical treatments of core-collapse
supernovae, which suggest that neutron stars could plausibly be born
with spin periods of 1 ms and $a \simeq 0.7$ (e.g. Heger, Langer, \&
Woosley 2000).  A variety of different effects have been proposed to
explain how a neutron star might be rapidly spun-down to a point that
agrees with inferred natal spin periods.

R-mode instabilities on the surface of a hot neutron star might lead
to significant gravitational wave losses, and quickly increase the
spin period (Andersson 1998, Andersson, Kokkotas, \& Schutz 1999).
Due to the low saturation amplitude of the instabilities and their
strong spin-frequency dependence, however, it may take thousands of
years to spin down neutron stars with moderate initial spins (Arras et
al.\ 2003, Heger, Woosley, \& Spruit 2005, Ott et al.\ 2006).
Advances in gravitational wave detection may eventually enable a test
of this spin-down mechanism.

Spin-down via the magnetic ``propeller'' mechanism may
be another viable means of quickly increasing the period of neutron
stars that are born spinning very rapidly.  However, this picture
requires significant ``fallback'' accretion after the initial
explosion, and fallback may not be important in the majority of
supernovae (Ott et al.\ 2006).

Spruit \& Phinney (1998) have proposed that neutron star spins might
be imparted by off-center kicks rather than angular momentum
conservation within a contracting star.  In this picture, the
progenitor star rotates as a solid body prior to collapse, leaving a
very slowly-rotating neutron star.  Asymmetries in the collapse give
an off-center kick to the neutron star, imparting the higher spins
that are observed.  Thus, pulsar kicks and spin periods should be
positively correlated.  This is a particularly appealing possibility
in that two phenomena are explained through a common and fairly simple
process.  However, the predicted link between spin period and kick
velocity may be at odds with data (see, e.g., Kaspi \& Helfand 2002).

Differential rotation in a massive star may be a viable means of
producing the observed distribution of natal neutron star spins, and
obviate the need for spin-down mechanisms that act shortly after
birth.  Heger, Woosley, \& Spruit (2005) examine the effects of
differential rotation in a $15~{\rm M}_{\odot}$ star, and find that
plausible magnetic torques could reduce the rotation rate of the
neutrons star by factors of 30--50 compared to scenarios where
magnetic fields are unimportant.  The specific angular momentum of the
pre-collapse core predicted in these magnetic torque models is {\it
  below} the threshold required to produce strong MHD jets in the
``collapsar'' model of MacFadyen \& Woosley (1999), for enclosed
masses less than $3.2~{\rm M}_{\odot}$, and perhaps even up to $6~{\rm
  M}_{\odot}$.  This may imply that black holes with small masses
could have relatively low spin parameters.

At least within the framework of prevailing models for core-collapse
supernovae, and subject to the limitations and systematics of the
small, new samples examined, our results are consistent with most
black holes being born with relatively high spin in GRB-like
supernovae events with jets, while neutron stars (and a small subset
of black holes) may be born in core-collapse supernovae wherein some
effect -- plausibly magnetic torques -- has acted to reduce the
angular momentum of the stellar core and thus inhibited jet
production.  It is unclear why magnetic torques, fallback, or other
effects may be manifest in some core-collapse events, but not others.

As the number of black hole spin measurements and natal neutron star
spin periods increase over time, and as the field matures, the results
of this early analysis may change to a degree.  A more enduring result
may simply be that we are now able to utilize black hole angular
momenta, natal neutron star angular momenta, and comparisons between
the two as tools to better understand core-collapse supernovae and massive
stellar evolution.

\vspace{0.2in}
We thank Charles Gammie, Andy Fabian, Victoria Kaspi, and Stuart
Shapiro for helpful discussions.  We thank the referee, David Helfand,
for a careful review that improved this manuscript.  MCM acknowledges
support through NSF grant AST0708424.  CSR acknowledges NASA support
under grant NNX09AC09G.


\end{document}